\documentstyle[gcdv,psfig]{article}
\begin{document}

\small
\shorttitle{The Fifth Workshop on Galactic Chemodynamics}
\shortauthor{B.K. Gibson and D. Kawata}

\title
{\large \bf
The Fifth Workshop on Galactic Chemodynamics
}

\author{\small 
 Brad K. Gibson$^12,$ \& Daisuke Kawata$^1$
}

\date{}
\twocolumn[
\maketitle
\vspace{-20pt}
\small
{\center
$^1$Centre for Astrophysics \& Supercomputing, 
Swinburne University, Hawthorn VIC 3122, Australia\\
$^1$bgibson,dkawata@astro.swin.edu.au\\[3mm]
}
\vspace{.7cm}
]

\begin{table}
\caption{
List of participants, home institutions, and manuscripts
}
\begin{center}
\begin{tabular}{lll}
  \hline
Johannes Andersen & Copenhagen & p.129 \\
Dominik Argast &  Basel &  p.161 \\
Tim Beers &  Michigan State & p.207 \\ 
Kenji Bekki & UNSW & p.167 \\ 
Joss Bland-Hawthorn &  AAO & p.110\\ 
Masashi Chiba &  Tohoku & p.237 \\ 
Chris Brook &  Swinburne & p.153 \\
David Burstein &  Arizona State &  \\
Bruce Carney &  North Carolina & p.134 \\
Tim Connors &  Swinburne & p.222 \\
Lisa Elliott &  Monash &  \\
Chris Flynn &  Tuorla & pp 126, 153\\
Ken Freeman &  ANU & p.110 \\
Ortwin Gerhard &  Basel & \\
Brad Gibson &  Swinburne & pp 153, 216, 222 \\
Stefan Harfst &  Kiel & p.228 \\
Amina Helmi &  Groningen & p.212\\
Gerhard Hensler &  Vienna & pp 188, 228 \\
Janne Holopainen & Tuorla & p. 153 \\
Akihiko Ibukiyama &  NAOJ & p.121\\
Inese Ivans &  Caltech & \\
Daisuke Kawata &  Swinburne & pp 153, 222 \\
Alexander Knebe &  Swinburne & p.216 \\
Chiaki Kobayashi &  MPA & p.183 \\
Hyun-chul Lee &  Swinburne & p.153 \\
Geraint Lewis &  Sydney & p.203 \\
Steve Majewski &  Virginia & p.197 \\
Masao Mori &  Senshu & p.232 \\
Naohito Nakasato &  Tokyo & p.171 \\
Julio Navarro &  Victoria & \\
Birgitta Nordstr\"om &  Lund & p.129 \\ 
Laura Portinari &  TAC & p.144 \\
Simone Recchi &  MPA & p.157 \\
Agostino Renda &  Swinburne & p.153 \\
C\'eline Reyl\'e &  Besancon & p.138 \\
Markus Samland &  Basel & pp 161, 175 \\
Arnaud Siebert &  Arizona & \\
Rainer Spurzem &  Heidelberg & p.188 \\
Takeru Suzuki &  Kyoto & p.148  \\
Matthias Steinmetz &  Potsdam & \\
Christian Theis &  Vienna & pp 179, 188, 228 \\
Chris Thom &  Swinburne & \\
Patricia Tissera &  IAFE & p.192\\
Eline Tolstoy &  Groningen & \\
Takuji Tsujimoto &  NAOJ & p.242 \\
Hideki Yahagi &  NAOJ & \\
  \hline
 \end{tabular}
\end{center}
\end{table}

From 1992 to 1995, the annual Workshops on Galactic Chemodynamics 
were recognised as
{\it the} primary cross-disciplinary meetings for theorists and observers
interested in understanding the formation and evolution of galaxies - 
in particular, the Milky Way.\footnote{The locations for the first 
four Workshops on Galactic Chemodynamics
were Clemson (1992), Kiel (1993), Livermore (1994), and Ringberg (1995).}
The eight years subsequent to the Fourth Workshop have seen an 
extraordinary expansion in the field of chemodynamics, driven not only
by the obvious advances in computational power, but by an incredible
wealth of new observational data --- ranging from the discovery of 
multiple chemo-kinematical substructures in the Milky Way and M31 halos, to
the discovery of the 
most metal-poor object known in the Universe (also within the halo
of the Milky Way).  Further, the recognition that ``near-field cosmology''
--- deconstructing the formation and evolution history of our Milky 
Way on a star-by-star basis -- was the primary science driver
for ambitious next-generation surveys and facilities such as the
RAdial Velocity Experiment 
(RAVE)\footnote{\tt http://astronomy.swin.edu.au/RAVE/} 
and the European Space Agency's
${\it Gaia}$ mission\footnote{\tt http://astro.estec.esa.nl/GAIA/}, led us to
believe that a Fifth Workshop on Galactic Chemodynamics (GCDV)
was long overdue.

In September 2002, we put out a call for participation in GCDV and were
overwhelmed with the number of positive responses.  Our goal with GCDV was
to once again bring together the leaders in the computational and
observational fields, in order to galvanise efforts related to 
deconstructing the history of formation of the Milky Way.

GCDV took place 9$-$11 July 2003, at Swinburne University, Melbourne, 
Australia.  A remarkable 46 participants from 13 countries attended 
the Workshop; 28 of the 36 talks presented there have been included in these 
proceedings.

The broad topics covered during the meeting included:
\begin{itemize}
\item the formation of the Milky Way in a Cold Dark Matter universe: 
merging versus smooth accretion
\item the connection between the halo, bulge, 
and thick + thin disk components
\item correlations between chemical and dynamical properties of stars
in the Milky Way
\item the (homogeneous and/or inhomogeneous) chemical enrichment history
of galaxies
\item evolution of the Galactic multi-phase interstellar medium
\item self-regulating star formation and chemodynamics
\item computational methods of galaxy evolution (Nbody, Smoothed Particle 
Hydrodynamics, Adaptive Mesh Refinement)
\end{itemize}

The Scientific Organising Committee was charged with ensuring that the
above
ambitious science scope was met, and as the 28 papers presented
in these proceedings demonstrate, they appear to have succeeded.  The SOC
was Chaired by Brad Gibson (Swinburne), and supported ably by
Andi Burkert (Munich), Gerhard Hensler (Vienna), and Daisuke Kawata
(Swinburne).

The success of GCDV can be traced in no small part to the support 
provided by the Local Organising Committee.  While Chaired by Brad
Gibson and Daisuke Kawata, it was really the efforts of Michelle Jolley,
Chris Brook, and Chris Thom that allowed the Workshop to function
seamlessly.

The publication of the proceedings of GCDV was made possible through the
the support of the \it Publications of the
Astronomical Society of Australia\rm.  We especially wish to
acknowledge the tireless efforts of its editors Louise Hartley and 
Richard Hecker.  Further, while Brad Gibson and 
Daisuke Kawata oversaw
the editorial duties for these particular proceedings, their editorial
assistants --- Chris Brook, 
Yeshe Fenner, Agostino Renda, and Chris Thom 
--- deserve
special mention for chasing down authors and referee reports for each
of the manuscripts.

\newpage

Finally, we would like to express our thanks to each of the Workshop
participants.  The overwhelming response to the Call for a Fifth Workshop
on Galactic Chemodynamics (we had anticipated fewer than 20 interested
participants!) demonstrates the need for the community to 
gather for such targeted meetings on a more regular basis.  We 
can only hope that the primary legacy of GCDV will be the 
re-establishment of
an annual (or at least biennial) Galactic Chemodynamics Workshop series ---
it is clear that the demand and desire exists.

\bigskip
\bigskip

\hfill
Brad K. Gibson

\hfill
Daisuke Kawata

\hfill
June 2004

\end{document}